# Locally Polarized Wave Propagation through Metamaterials' Crystallinity

Simon Yves[1], Thomas Berthelot[2,3], Geoffroy Lerosey[4] and Fabrice Lemoult[1]

[1] *Institut Langevin, CNRS UMR 7587, ESPCI Paris, PSL Research University,*
*1 rue Jussieu, 75005 Paris, France*
[2] *NIMBE, CEA, CNRS Université Paris-Saclay, CEA Saclay, 91191 Gif sur Yvette Cedex, France*
[3] *KELENN Technology, Antony, France*
[4] *Greenerwave, ESPCI Paris Incubator PC'up, 6 rue Jean Calvin, 75005 Paris, France*
*\*Correspondence to [fabrice.lemoult@espci.fr](fabrice.lemoult@espci.fr)*

Abstract

Wave propagation control is of fundamental interest in many areas of Physics. It can be achieved with wavelength-scaled photonic crystals, hence avoiding low frequency applications. By contrast, metamaterials are structured on a deep-subwavelength scale, and therefore usually described through homogenization, neglecting the unit-cell structuration. Here, we show with microwaves that, by considering their inherent crystallinity, we can induce wave propagation carrying angular momenta within a subwavelength-scaled collection of wires. Then, inspired by the Quantum Valley-Hall Effect in condensed matter physics, we exploit this bulk circular polarization to create modes propagating along particular interfaces. The latter also carry an edge angular momentum whose conservation during the propagation allows wave routing by design in specific directions. This experimental study not only evidences that crystalline metamaterials are a straightforward tabletop platform to emulate exciting solid-state physics phenomena at the macroscopic scale, but it also opens the door to crystalline polarized subwavelength waveguides.

# Main

In Nature, the interplay between light and the constituents of a material gives birth to interesting phenomena that have different origins. In the case of X-ray diffraction, bright spots, which are due to the constructive interferences of the waves reflected off consecutive Bragg planes, allow to determine the structure of the crystal[1]. As for infra-red or ultraviolet spectroscopies, the chemical sample under consideration, via molecular resonances, generates absorption's peaks in the spectrum, which provides information about the composition of the medium [2]. From these observations, a paradigmatic change emerges which consists of designing the structure and the composition of materials in order to tailor the wave propagation. This is the domain of metamaterials[3,4,5], that can be seen as the macroscopic analogues of dielectrics[6], and consist in arrays of inclusions that are structured on a mesoscopic scale, smaller than the incoming wavelength. Hence, new properties have been demonstrated such as negative refraction[7,8], cloaking [9,10], zero permittivity[11] or huge refractive index[12,13] for example.

We now focus on a particular class of these composite media, namely the locally resonant ones, whose unit cell is a subwavelength resonator. Without loss of generality, we work in the microwave domain and take quarter-wavelength copper wires on a ground plane as resonant units. Recent studies have shown that multiple scattering exists within these systems, albeit the deep-subwavelength distances at stake [14,15,16]. In particular, a metamaterial arranged as a honeycomb lattice, as displayed on Fig.1(a), exhibits a propagating band for which negative refraction occurs 14. Moreover, the band structure (Fig.1(c)) presents single point degeneracies at the different corners $K$ and $K'$ of the Brillouin zone. These are known as Dirac cones, also named Valleys, and rely on the honeycomb pattern. In a previous work, we demonstrated that these degeneracies can be lifted by inducing small structural deformations, which creates a pseudo-spin degree of freedom in the medium[17]. This leads to a macroscopic analog of a $Z_2$ topological insulator which present robust wave propagation on their edges [18,19,20,21].

In this letter, we lift this degeneracy by controlling the composition of the medium: we simply detune one of the two resonators. This not only opens a bandgap, but also induces a microscopic angular momentum. Following the example of the Valley-Hall topological insulator in condensed matter physics [22], we experimentally demonstrate the guiding of waves along specific interfaces. These edge-modes also possess a crystalline polarization that is straightforwardly described in terms of angular momentum. The conservation of this additional degree of freedom allows then to build directional junctions, thus routing the microwaves in specific directions. Moreover, these guided modes being localized on the interface, we experimentally show that the previous results hold if we only keep the interface's resonators that carry the angular momentum, and replace the remaining crystal by a randomly patterned collection of identically detuned wires.

As introduced before, we start from a metamaterial crystal which consists in a honeycomb lattice of copper rods of height $h_0 = 13.9$ mm standing on a ground plane (Fig.1(a)). The resonance frequency of these "meta-atoms" is inversely proportional to their length. The distance between each neighboring wire is $5$ mm, far smaller than the free-space wavelength ($\lambda_0 \approx 6$ cm). The corresponding dispersion relation (Fig.1(c)), obtained by simulating the unit cell and applying Bloch periodic boundary conditions in COMSOL, presents two bands that come from the folding of a polariton [14,15]. Moreover, at the different corners of the Brillouin zone, $K$ and $K'$, the dispersion relation exhibits the previously mentioned Dirac cones.

To lift these degeneracies, we play with the composition of the subwavelength unit cell and we slightly detune the resonators (Fig.1(b)). This is done straightforwardly by taking a shorter wire (white, $h_2 = 13.8$ mm) that resonates at a higher frequency and a longer one (black, $h_1 = 14$ mm) which resonates at a lower frequency. The newly simulated band structure (Fig.1(d)) exhibits the disappearance of the Dirac cones. The numerically calculated electric field distributions corresponding to the two bands at the corners $K$ and $K'$ are presented on Fig.1(e) to (h). To guide the reader, we define a fixed hexagon delimited by the black contour on the field maps: it is composed of two triplets of resonators with different resonance frequencies displayed as triangles which either have their basis at the bottom (defined as green), or at the top (defined as yellow) with respect to the $\vec{K}$ Bloch wavevector.

First, we focus on the lower band at the $K$ corner (Fig.1(g)). The field is only concentrated on the black wires, which indeed resonate at the lowest frequency, and is null on the white ones. Also at this $|\vec{K}|$ wavenumber, the electric field oscillates with a phase delay of $2\pi/3$ between each black wire site: if the field is maximum on one summit of the triangle, it will be minimum on another one and null on the third one. Sweeping the phase to mimic the propagation within the infinite crystal points out the existence of a rotation of the electric field:

the green link in Fig.1(g) connects all of the field's maxima and results in a right-handed helix.

For the upper band (Fig.1(e)), the field is obviously distributed on the white wires that have a higher resonance frequency. The phase sweep shows again a rotation of the electric field within the hexagon, but it turns the other way compared to the lower band, as shown by the corresponding yellow left-handed helix that follows the field's maxima. In the case of a wave travelling with the $\overrightarrow{K'}$ wavevector, the field maps (Fig.1(f) and (h)) evidence a reversed behavior compared to the $K$ point. We note here that this is coherent with the symmetrical analysis of the crystal: albeit the loss of the inversion symmetry, the lattice still presents the time-reversal one.

In order to check the pertinence of this numerical work, an experimental sample is prepared by metallizing a 3D printed ABS-based plastic medium. The finite-sized sample (Fig.2(a)), whose total dimension is comparable to $\lambda_0$, has a triangular shape in order to respect the lattice's symmetries. A small loop antenna excites the sample in the close vicinity of an exterior wire, and a near-field probe connected to a network analyzer measures the electric field right above the wires. We carry out a scanning of the full area of the metamaterial and obtain field maps for a broad range of frequencies. Moreover, as we want to select one direction between $K$ and $K'$ to observe the local rotation of the electric field, we put the source slightly off the symmetrical axis of the triangle so as to favor one of these directions.

The averaged spectrum measured on the central area of the metamaterial (Fig.2(b)) is composed of peaks which traduce the existence of resonant eigenmodes because of the finiteness of the sample. These peaks can be gathered in two groups: one below $4.98$ GHz and one above $4.92$ GHz, separated by a bandgap (shaded area). The field maps corresponding to the peaks just below and above the bandgap are represented respectively in Fig.2(c) and (e). We again stress that everything happens at a deep subwavelength scale, highlighted by the numbers of sign change within roughly one free-space wavelength. The Fourier maps corresponding to those measurements (Fig.2(d) and (f)) have a non-zero amplitude only at the Brillouin zone's corners corresponding to a single Valley.

We then carry out the same procedure as for the previous numerical study: we focus on a hexagonal meta-molecule that we take at the middle of the sample (black line on the field maps). As expected, for the low frequency map the field is distributed on the longer black wires, while above the bandgap it is on the white shorter ones. We evidence the existence of the crystalline angular momentum by again sweeping the phase (Fig.2(h) and (g)). Similar helices are obtained, confirming experimentally the creation of an angular momentum within the bulk crystalline metamaterial. Also, it is left-handed below the gap while it is right-handed above, in agreement with the numerical simulations.

This particular rotational polarization of the electric field, which is linked to the $K$ or $K'$ direction, is actually a macroscopic manifestation of the Valley degree of freedom in solid state physics [22]. Indeed, in graphene-like structures, the electronic wave-function undergoes a crystalline circular polarization whose sense of rotation depends on the Valley. The latter produces an intrinsic magnetic moment comparable to the spin of the electron[23] which can similarly be linked to a specific topological invariant, namely the Valley Chern number. Hence, an interface between two crystals which is designed to put different Valleys one opposite to the other is able to carry propagating modes in the bandgap. These modes present an additional degree of freedom, namely the Valley polarization, which limits backscattering and can be used to route selectively the waves. This is known as the Quantum Valley-Hall effect and has been demonstrated not only in electronic structures [24,25] but also at a macroscopic scale with classical waves [26,27,28,29,30,31,32,33].

Consequently, after having demonstrated the existence of this Valley degree of freedom in the bulk of the subwavelength crystal, we need to design an interface between two metamaterials. We have to go to the reciprocal space and find a way to put the $K$ point in front of $K'$ at the domain wall between the two crystals. A simple way to achieve this is to add a mirror symmetry along the $KK'$ axis. It leads to two kinds of interfaces between a crystal and its symmetrical one: the first interface (Fig.3(a)) results in putting two long wires (black) next to each other, while the other (Fig.3(e)) does the same with the short wires (white). The corresponding simulated dispersion relations for each interface are displayed in Fig.3(b) and (f).

Let us focus on the first case. A propagating band (red line) exists in the bandgap of the bulk modes (dark shaded area). A closer look at the electric field map which corresponds to a mode propagating towards the right (Fig.3(c)) reveals that the field is mainly localized on the resonators at the interface, being confined laterally on a deep subwavelength scale. As the field is localized only on the wires of the domain wall, we want to describe the wave propagation locally in the real space in the same manner as what we did before in the case of the bulk. Hence, we define a hexagon-shaped meta-molecule made of six resonators of the interface between the two crystals (black line on Fig.3(c)). The latter defines again two triangles with different orientations with regard to the $x$-direction. Coherently with the colours of Fig.1, they are displayed in green and yellow on the sketch of Fig.3(d). However, although the structure of this meta-molecule is similar, its composition is not the same. Indeed, the two triangles are not made of a single type of resonators (black or white) but both of them possess two low resonance frequency black ones and one high resonance frequency white one. And, the electric field is now distributed on both of those two hybrid triangles. Following the maxima of the field separately on each of the triangles while sweeping the phase results in two different spirals as pictured on Fig.3(d): the yellow one is left-handed while the green one is right-handed. Hence, this edge mode which propagates along the red interface presents a combination of two counter-rotating circular polarizations which are locked to the direction of propagation. This superposition results in this case in an anti-symmetric mode at the interface.

For the other type of interface (Fig.3(e)), the numerical dispersion relation of Fig.3(f) also exhibits a band in the bandgap frequency range of the bulk (blue line). However, in this case, because the mode which propagates to the right must have a positive group velocity, the pertinent field map corresponds to a negative $k_x$ value. The latter (Fig.3(g)) shows a symmetric profile and this particular symmetry presents dipolar modes along the propagation which is coherent with the negative slope of this band. The same microscopic study as before results in a different meta-molecule: the white short wires and the black long wires are exchanged compared to the previous case. Finally, we show in Fig.3(h), that the propagating wave can also be characterized by a combination of two counter-rotating helices locked to the propagation direction that are reversed compared to the ones of the red interface.

This numerical study shows that, by making an analogue of the Quantum Valley-Hall effect in a crystalline metamaterial, we are able to guide microwaves on a subwavelength scale at the interface between two crystals. Moreover, adopting a microscopic description of the guided modes at the level of the interface, we clearly evidence the retrieval of a combination of angular momenta locked to the propagation direction. We also show that these mixed circular polarizations rotate in opposite sense according to the type of domain wall, or in other words, according to the composition of the meta-molecule at the interface.

The next step is to characterize experimentally this peculiar property of the guided waves. We start by making a sample which exhibits a blue type domain wall with subwavelength sharp turns between two metamaterial crystals (Fig.4(a)). In order to highlight the propagation properties of the guided modes we work in the temporal domain and we carry out the following experiments: a small loop antenna placed in the close vicinity at the bottom of the sample (position 1) emits a 20-ns-pulse centered at the central frequency of the Valleys (4,95 GHz). The intensity over time is measured on the domain wall at the middle of the sample after the first sharp turn (position 2) and at the end of the bended interface (position 3). The transient transmissions presented Fig.4(a) show that the pulse propagates along the path with very little backscattering. Snapshots of the pulse intensity at two times $t_1$ and $t_2$, corresponding to the moments where the pulse successfully takes the first and second sharp turn respectively, presented in Fig.4(b), distinctly evidence the robust propagation of microwaves on a subwavelength scaled distorted path. It also shows a symmetric profile of the mode which is relevant with the one described on Fig.2(g). Interestingly, we can relate this singular behavior to the local field rotation at the level of the interface. This is depicted in Fig.4(c), where the blue interface is sketched using the meta-molecule presented previously in Fig.3(h). Here, for the sake of clarity we only represent the rotation on the green triangle, but one should keep in mind that the field is also rotating on the inverted yellow triangle in the opposite sense. This picture shows markedly that the specific bends of the interface, dictated by the reciprocal crystalline arrangement of the sample, always preserve the sense of rotation on the interface. Hence, this local microscopic description explains genuinely the absence of backscattering of the guided waves.

A more complex case (Fig.4(d)) consists in a six branches crossing made of red interfaces. We again send a 20-ns-pulse in the branch number one and we measure the transmission at the end of the other branches. The wave-packet only propagates in the branches 2 and 6, and in branch 4 with a small delay, but nothing goes in the 3 and 5 pathways. Even though ohmic losses are small in the medium, the reduction of the pulse intensity should be accounted for dissipation during the propagation. A snapshot of the pulse intensity taken at time $t_3$ (Fig.4(e)) confirms the expected anti-symmetric nature of the propagating mode, as well as proves the specific routing of the energy within the sample. The latter is again explained in a simple way by the locking of the local sense of rotation on the interface with the direction of propagation (Fig.4(f)).

After the separate study of the blue and the red interfaces, we carry out experiments on a sample which mixes the two types of domain walls (Fig.4(g)). In this case, the pulse, still starting from the position 1, travels to the positions 2 and 4 but does not go to 3, as it is also shown on the snapshot taken at time $t_4$ (Fig.4(h)). Actually, in that case we take advantage of another property of these guided modes which is that the locking of the angular momentum with the direction has an opposite sense of rotation on being on the red interface meta-molecule or on the blue one. This consequently explains the experimental results as sketched on Fig.4(i).

This set of experiments in the temporal domain makes a strong evidence of the role of the circular polarization locking with the propagation's direction on the different interfaces. In the solid-state physics domain, this behavior is known as Valley conservation. However this Valley appellation refers to the band structure of the crystal, hence to a global property of an infinite structure and furthermore in the reciprocal domain. Here we explain all the experimental results simply by adopting a local description of the electric field behaviour at the interface, in the real space. We believe that this microscopic approach gives a genuine insight of the physics at play here.

Another consequence of this approach is that it reveals the role of the wires far from the interface: they only play the role of preventing the leakages of the edge modes in the bulk thanks to the opened bandgap. However, we actually know a much simpler way to generate a highly controllable bandgap in the context of locally resonant metamaterials. Because of the resonant nature of the unit cell, the wave physics within such a system is of polaritonic nature [34]. This polariton exhibits a hybridization bandgap above the resonance of a single resonator which only relies on the resonance property [35]. This bandgap can be used to induce defects straightforwardly in the medium, the latter being inclusions with a higher resonance frequency which fall in the previous bandgap frequency range. Its high versatility has been used in recent studies to create subwavelength cavities [35], ultra-compact waveguides [36] and delay lines [37].

Following these ideas, we make samples similar to the ones presented in Fig.4(d) and (g) but now we keep only the interfaces' wires and replace all of the remaining crystal by a spatially random collection of longer green wires of height $h3 = 16$ mm (Fig.5(a) and (e)). Because these green wires have a lower resonance frequency compared to the black and white ones, the modes supported by the red and blue interfaces fall into the bandgap frequency range of the surrounding disordered media. We hence created quasi-one dimensional defects waveguides and carried identical temporal experiments as before. In the case of the six red branches crossroads (Fig.5(b)) the same routing of the pulse as in the previous experimental study is demonstrated: only the branches 2, 4 and 6 present signal. Actually a small amount of intensity exists for longer times in the branch 3, around 40 ns. This comes from the reflection of the wave-packet at the end of the waveguides, nothing being impedance matched in this study. If we look at the snapshot of the pulse intensity taken at time $t_1$ (Fig.5(c)), we clearly see the specific routing of the microwaves thanks to the local rotational locking of the propagation. Moreover, the bandgap properties here are different than the previous crystalline one and the electric field is far more confined on the interface resonators. As for the sample which mixes the blue and red interfaces, the results (Fig.5(f) and (g)) again distinctly show the same propagation properties as before.

In this paper, we markedly evidence the great potential of the bottom-up description of locally resonant metamaterials. Particularly, we show that slightly tuning the composition of a structured subwavelength scaled honeycomb lattice not only leads to a bandgap opening, but also to the creation of a crystalline rotational moment, defined on a bulk meta-molecule, which is locked to the propagation direction of the waves within the crystal. We then take advantage of this bulk property in order to create a macroscopic classical analogue of the Quantum Valley-Hall effect which leads to Valley-polarized guided modes. Although this is precisely described by theories based on the definition of topological invariants, here we deliberately intended to give a microscopic description of this phenomenon in the real space. We demonstrated that an external crystalline degree of freedom, that is a combination of two counter-rotating angular momenta confined on the interface' resonators, is locked to the propagation direction of these guided modes. These newly defined meta-molecules emerge from the hybridization of the bulk ones and their properties rely on their composition. Finally, we proved that the same meta-molecules keeps this supplemental degree of freedom and behaves similarly if introduced in a spatially random collection of scatterers, which are here solely to create a hybridization bandgap.

To conclude, not only this study strongly demonstrates the great potential of crystalline metamaterials as a really genuine macroscopic tabletop platform to investigate tantalizing solid-state physics phenomena, but it also opens the door to applications of metamaterial for wave manipulation at the subwavelength scale.

**Acknowledgements**

S.Y. acknowledges funding from French Direction Générale de l'Armement. This work is supported by LABEX WIFI (Laboratory of Excellence within the French Program "Investments for the Future" ) under references ANR-10-LABX-24 and ANR-10-IDEX-0001-02 PSL* and by Agence Nationale de la Recherche under reference ANR-16-CE31-0015.

**Contributions**

SY performed the experiments. S.Y and F.L carried out the numerical simulations. TB developed the sample fabrication procedure. FL and GL supervised the project. All authors contributed to the research work and participated to the redaction of the manuscript.

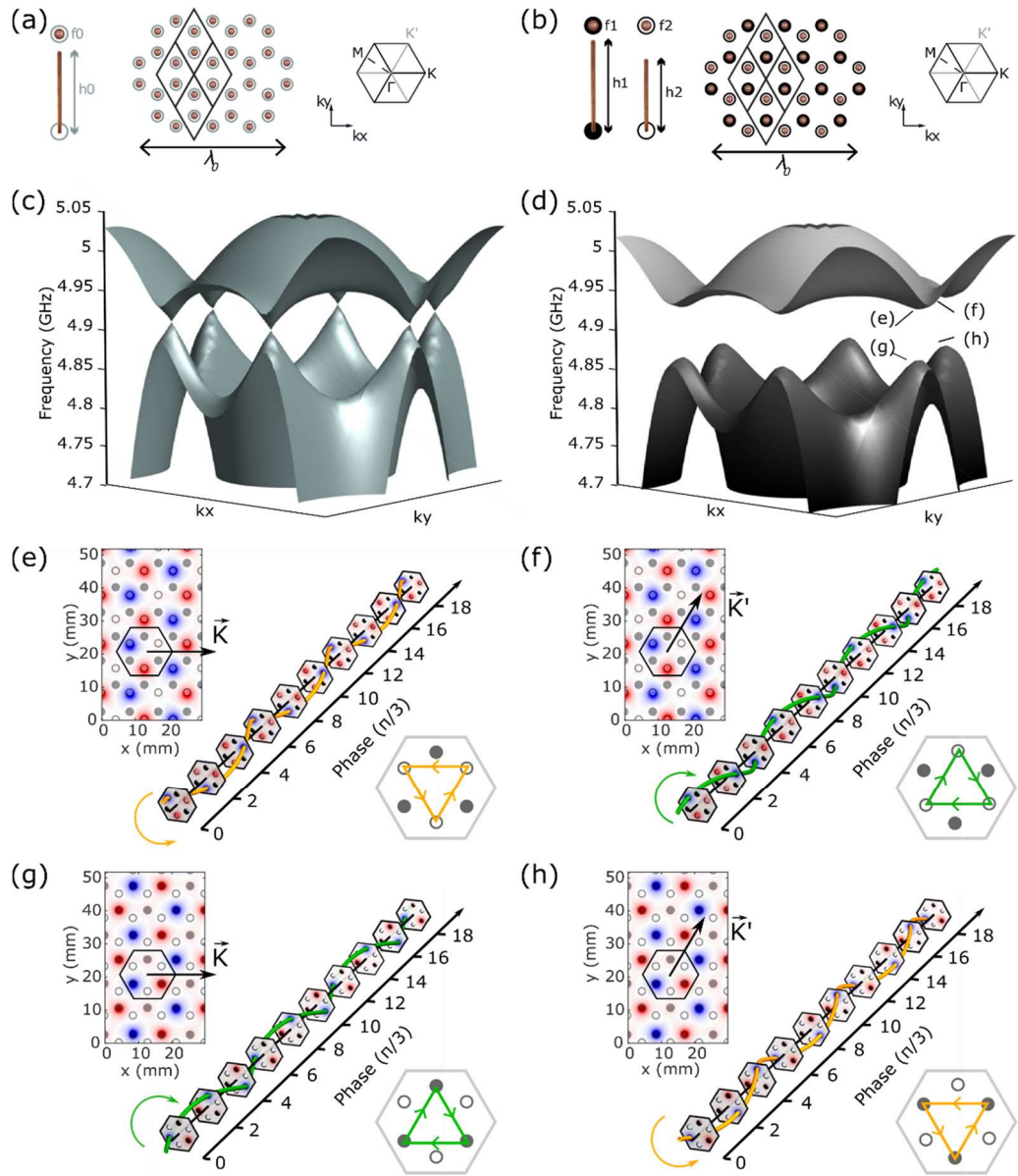

**Figure 1. Subwavelength band structure engineering to induce crystalline angular orbital momentum. (a)**.Crystalline metamaterial made of identical copper wires arranged as a subwavelength honeycomb lattice and its corresponding first Brillouin zone. **(b)** Same as **(a)** but with two different wires in the unit cell. **(c)** Subwavelength scaled band structure of **(a)**. **(d)** Same as **(c)** for **(b)**. **(e)** (resp. **(g)**) Numerical electric field map of the crystalline eigenmode of the upper (resp. lower) band at the $K$ corner of the Brillouin zone. In inset the corresponding bulk meta-molecules made of two triangles with either the basis at the bottom (green) or at the top (yellow) with regard to the wavevector. Demonstration of the different orbital angular momentum within the meta-molecule. **(f)** (resp. **(g)**) same as **(e)** (resp. **(g)**) but at the $K'$ corner of the Brillouin zone.

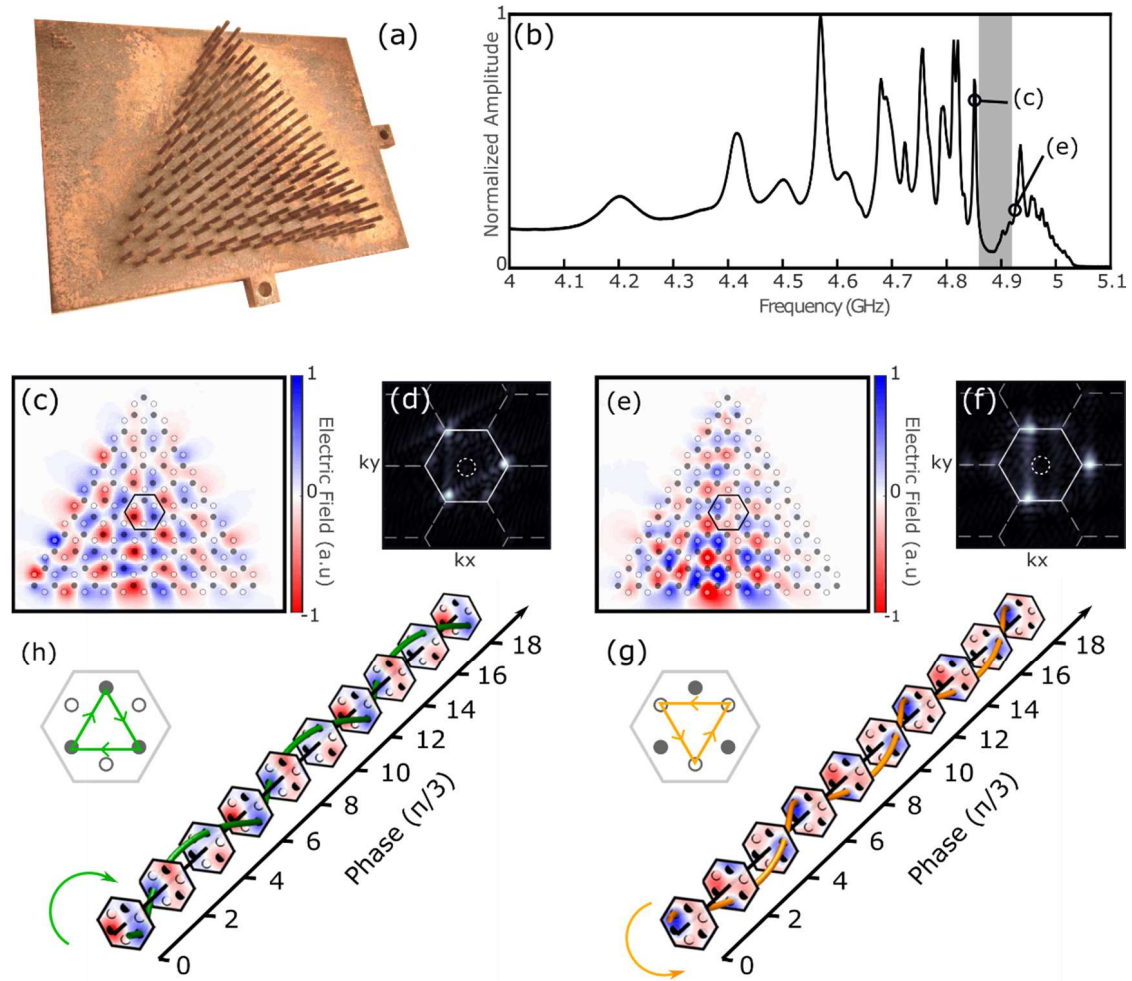

**Figure 2. Experimental measurement of the bulk crystalline orbital angular momentum. (a)** Picture of the sample. **(b)** Spectrum averaged on the center of the sample. It exhibits transmission peaks and a bandgap (shaded area). **(c)** (resp. **(e)**). Experimental electric field map of the mode below (resp. above) the bandgap. **(d)** (resp. **(f)**) Bi-dimensional Fourier transform of **(c)** (resp. **(e)**). Light cone is shown in dotted line. **(h)** (resp. **(g)**) Experimental demonstration of the existence of the crystalline orbital angular momentum of **(c)** (resp. **(e)**).

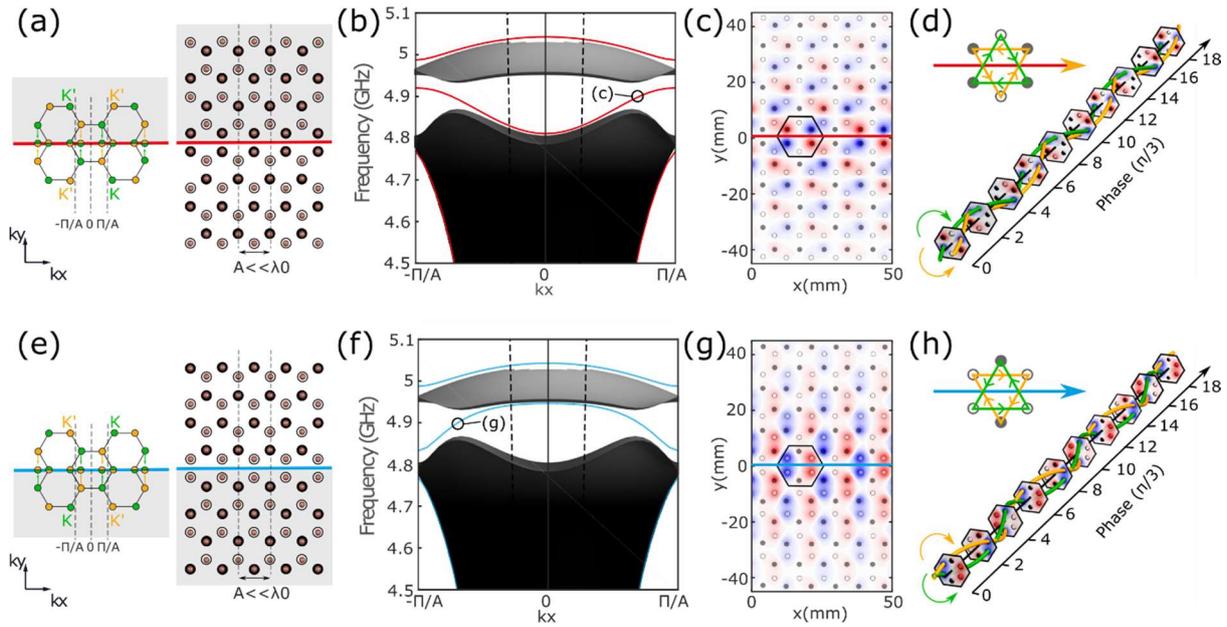

**Figure 3 Quantum Valley-Hall effect in a crystalline metamaterial. (a)** Red interface in the reciprocal and real space which corresponds to Valley overlapping. **(b)** Corresponding dispersion relation. **(c)** Electric field map of the red interface mode. **(d)** Corresponding meta-molecule with crystalline orbital angular momentum combination which is locked to the propagation. **(e), (f), (g), (h)** same as **(a), (b), (c), (d)** for the blue type interface.

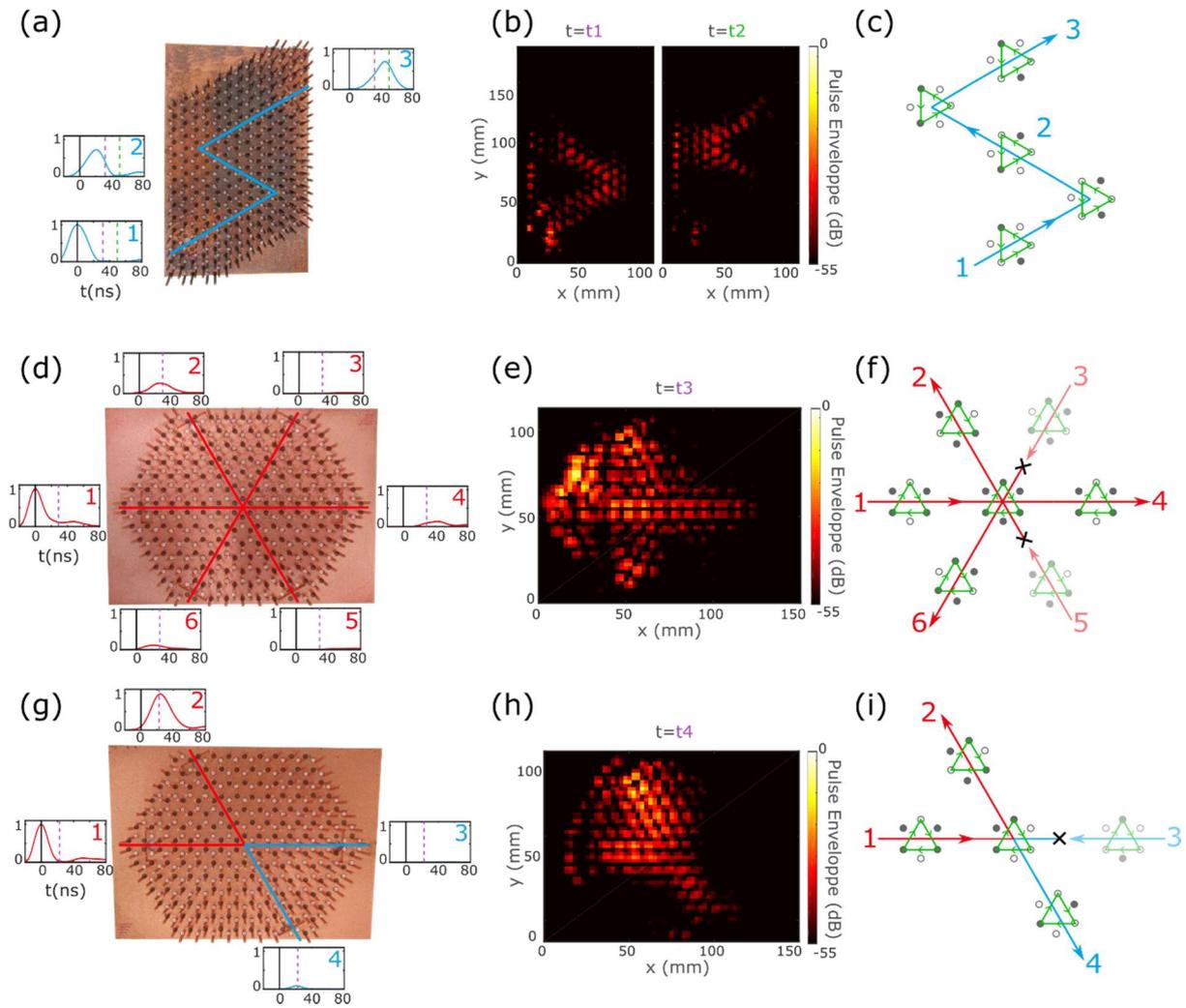

**Figure 4. Experimental demonstration of orbital angular momentum conservation. (a)** Sample and pulse propagation results in the case of a blue interface with sharp turns. **(b)** Snapshots of the pulse propagation after each bend. **(c)** Corresponding microscopic explanation with blue interface meta-molecules. **(d)** Sample and propagation results in the case of a crossroad made of six red interfaces. **(e)** Snapshot of the propagating pulse. **(f)** Corresponding microscopic explanation with red interface meta-molecules. **(g)** Sample and propagation results in the case of a crossroad mixing red and blue interfaces. **(h)** Snapshot of the propagating pulse. **(i)** Corresponding microscopic explanation with both red and blue interface meta-molecules.

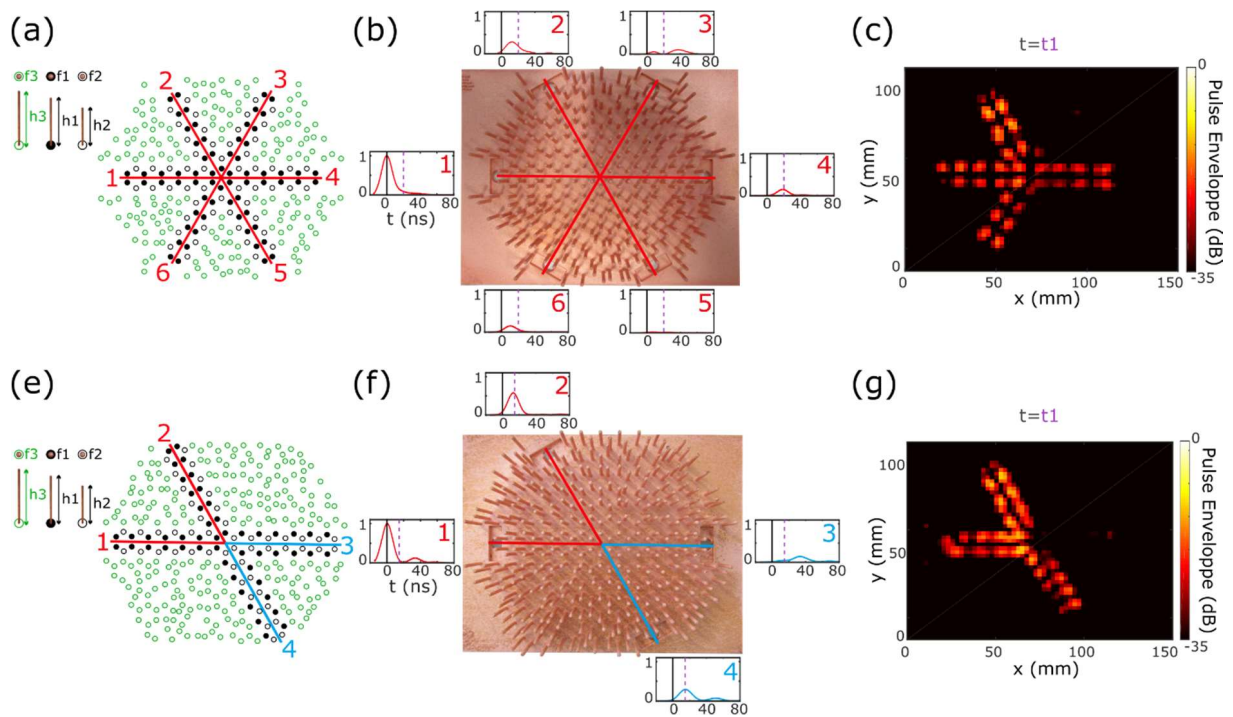

**Figure 5. Experimental subwavelength waveguides with crystalline polarization. (a)** Sketch of a crossroad which consist in six defect waveguides made of the red meta-molecule surrounded by a spatially random collection of low frequency resonators. **(b)** Sample and pulse propagation results corresponding to **(a)**. **(c)** Snapshot of the pulse propagation. **(e)** Sketch of a crossroad which consist in four defect waveguides made of both red and blue meta-molecule surrounded by a spatially random collection of low frequency resonators. **(f)** Sample and pulse propagation results corresponding to **(e)**. **(g)** Snapshot of the pulse propagation.